\documentclass[conference]{IEEEtran}
\IEEEoverridecommandlockouts
\usepackage{cite}
\usepackage{amsmath,amssymb,amsfonts}
\usepackage{algorithmic}
\usepackage{graphicx}
\usepackage{textcomp}
\usepackage{xcolor}
\usepackage{booktabs}
\usepackage{hyperref}
\def\BibTeX{{\rm B\kern-.05em{\sc i\kern-.025em b}\kern-.08em
    T\kern-.1667em\lower.7ex\hbox{E}\kern-.125emX}}
\begin{document}

\title{Wavelet Analysis of Noninvasive EEG Signals Discriminates Complex and Natural Grasp Types\\
\thanks{This project was supported by the Rhode Island INBRE program from the National Institute of General Medical Sciences of the National Institutes of Health under grant number P20GM103430 and National Science Foundation under award ID 2245558. This research was also supported by URI foundation grant on Medical Research. Author 5 is the corresponding author.}
}


\author{\IEEEauthorblockN{1\textsuperscript{st} Ali Rabiee}
\IEEEauthorblockA{\textit{Dept. of Electrical, Computer and Biomedical Engineering} \\
\textit{University of Rhode Island}\\
Kingston, RI, USA \\
ali.rabiee@uri.edu}
\and
\IEEEauthorblockN{2\textsuperscript{nd}  Sima Ghafoori}
\IEEEauthorblockA{\textit{Dept. of Electrical, Computer and Biomedical Engineering} \\
\textit{University of Rhode Island}\\
Kingston, RI, USA \\
sima.ghafoori@uri.edu}
\and
\IEEEauthorblockN{3\textsuperscript{rd} Anna Cetera}
\IEEEauthorblockA{\textit{Dept. of Electrical, Computer and Biomedical Engineering} \\
\textit{University of Rhode Island}\\
Kingston, RI, USA \\
annacetera@uri.edu}
\and
\IEEEauthorblockN{4\textsuperscript{th} Yalda Shahriari}
\IEEEauthorblockA{\textit{Dept. of Electrical, Computer and Biomedical Engineering} \\
\textit{University of Rhode Island}\\
Kingston, RI, USA \\
yalda\_shahriari@uri.edu\\
\IEEEmembership{Member, IEEE}}
\and
\IEEEauthorblockN{5\textsuperscript{th} Reza Abiri}
\IEEEauthorblockA{\textit{Dept. of Electrical, Computer and Biomedical Engineering} \\
\textit{University of Rhode Island}\\
Kingston, RI, USA \\
reza\_abiri@uri.edu\\
\IEEEmembership{Member, IEEE}}
}


\maketitle


\begin{abstract}
This research aims to decode hand grasps from Electroencephalograms (EEGs) for dexterous neuroprosthetic development and Brain-Computer Interface (BCI) applications, especially for patients with motor disorders. Particularly, it focuses on distinguishing two complex natural power and precision grasps in addition to a neutral condition as a no-movement condition using a new EEG-based BCI platform and wavelet signal processing. Wavelet analysis involved generating time-frequency and topographic maps from wavelet power coefficients. Then, by using machine learning techniques with novel wavelet features, we achieved high average accuracies: 85.16\% for multiclass, 95.37\% for No-Movement vs Power, 95.40\% for No-Movement vs Precision, and 88.07\% for Power vs Precision, demonstrating the effectiveness of these features in EEG-based grasp differentiation. In contrast to previous studies, a critical part of our study was permutation feature importance analysis, which highlighted key features for grasp classification. It revealed that the most crucial brain activities during grasping occur in the motor cortex, within the alpha and beta frequency bands. These insights demonstrate the potential of wavelet features in real-time neuroprosthetic technology and BCI applications.
\end{abstract}



\section{Introduction}
Grasping is a fundamental motor skill critical to daily human activities. The ability to decode these movements, particularly from neural signals like EEGs, has significant implications in the fields of assistive technology \cite{rupp2014brain}, neuroprosthetics \cite{grimm2016closed}, and rehabilitation \cite{park2013rehabilitation} for patients with motor impairments. Advances in EEG technology have made it possible to capture brain activity related to motor functions, offering new avenues for understanding and interfacing with the human motor system \cite{padfield2019eeg}. 

\begin{figure}[ht!] 
\centering
\includegraphics[width=3.5in]{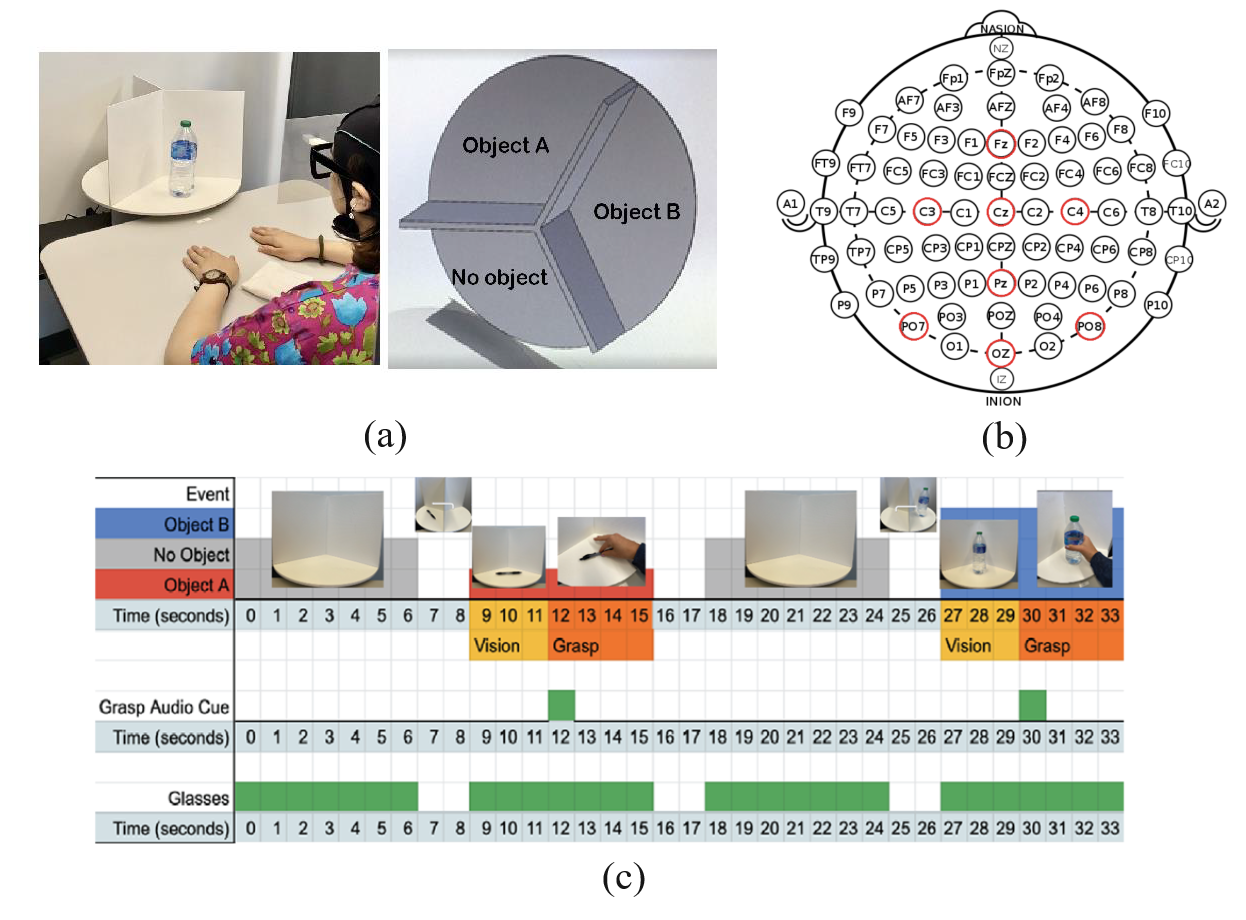}
\caption{Experimental setup and paradigm for reach-and-grasp tasks. (a) Platform components, (b) an illustration of a 128-channel EEG sensor layout with the 8 electrodes (in red) that are used in the study, and (c) The timeline of audio Cues and actions in the experimental paradigm.}
\label{experimental_paradigm}
\end{figure}

Wavelet signal processing has emerged as a powerful tool for analyzing non-stationary signals such as EEGs, offering a more nuanced understanding of brain activity in neurophysiology, particularly in decoding motor actions \cite{rhif2019wavelet}. This method enables the extraction of both time and frequency information from EEG data, which is crucial for identifying neural patterns associated with various types of hand grasps. Our study leverages the capabilities of wavelet analysis to address a critical research gap: while most existing studies concentrate on either time-domain or frequency-domain features \cite{xu2021decoding, schwarz2017decoding, schwarz2020analyzing}, there are limited studies that consider the integration of both. By combining these analyses, our research provides a comprehensive interpretation of EEG signals during complex motor tasks, enhancing the classification of different grasp types.

In this study, we utilized wavelet signal processing to analyze EEG data on a new BCI platform, focusing on classifying complex hand grasp types. We extracted novel wavelet-based features to identify unique neural patterns associated with power and precision grasps, as well as a no-movement condition. Using machine learning algorithms such as Support Vector Machine (SVM), Random Forest (RF), Extreme Gradient Boosting (XGBoost), and Linear Discriminant Analysis (LDA), we classified these patterns. A significant contribution of this research is the application of permutation feature importance analysis, which identifies the most critical features and electrode positions.


\section{Methods}

\subsection{Participants}

Five healthy participants, aged 23 to 35 and consisting of three females and two males, took part in the study. They provided written informed consent following a comprehensive briefing. The study received ethical clearance from the Institutional Review Board (IRB).

\subsection{Experimental Setup and Paradigm}

Our new experimental setup involved a 3D-designed motorized turntable partitioned into three sections, used for presenting two objects (a bottle and a pen) and a no-object condition. A key innovation in our platform development is using smart eyeglasses with an adjustable smart film that toggles between transparency and opacity, effectively eliminating any anticipatory bias. During the experiment, participants were seated in a comfortable chair in a neutral position, with palms facing down and positioned 30 cm away from the center of the object (as depicted in Figure 1a). As it is shown in Figure 1c after a 3-second observation phase, participants heard a buzzing sound, serving as an audio stimulus to initiate the grasping phase with their dominant hand. The turntable transitioned to the next object or no-object condition after each grasping phase, during which the smart eyeglasses were rendered opaque.

\subsection{Data Recording and Preprocessing}

EEG data were captured using the Unicorn Hybrid Black headset, a portable, wireless, dry-electrode EEG device equipped with eight recording electrodes and operating at a 250 Hz sampling rate (details available at \href{https://www.unicorn-bi.com}{Unicorn Hybrid Black headset}). The placement of the electrodes is depicted in Figure 1b. All recording electrodes were referenced to the mastoids, using linked mastoid referencing. We conducted 50 trials for each condition with every participant, thus compiling a comprehensive dataset for subsequent analysis.

The EEG data in the study were bandpass filtered between 1Hz and 30Hz to focus on frequencies relevant to grasping tasks. Artifacts were removed using Independent Component Analysis (ICA), followed by baseline correction using the 200 ms pre-event period. Finally, Z-score normalization was applied to standardize the signals across channels and participants, ensuring data integrity for analysis.

\subsection{Wavelet Analysis}

This study utilized complex Morlet wavelet analysis for EEG data, generating wavelets from 1 to 30 Hz across 4 frequency bins to examine detailed time-frequency information. The wavelets were defined by a sine wave within a Gaussian envelope, aiming for a balance between time and frequency precision \cite{cohen2019better}. With the number of cycles set to 4 and a 250 Hz sampling rate, these wavelets were convolved with EEG data, focusing on the power derived from the absolute value squared of the convolution. A key aspect of our analysis was plotting topographic maps of the wavelet power at specific time points, particularly 300 milliseconds after the movement onset, across selected frequencies (9, 16, and 23 Hz). These topographic maps provided spatial insights into the distribution of brain activity across different scalp regions. Additionally, we conducted a time-frequency analysis by plotting time-frequency maps. This analysis spanned from 1 second before to 1 second after the movement onset, offering a detailed view of the dynamic changes in brain activity across different frequencies over time. 

\begin{figure*}[ht!] 
\centering
\includegraphics[width=7 in]{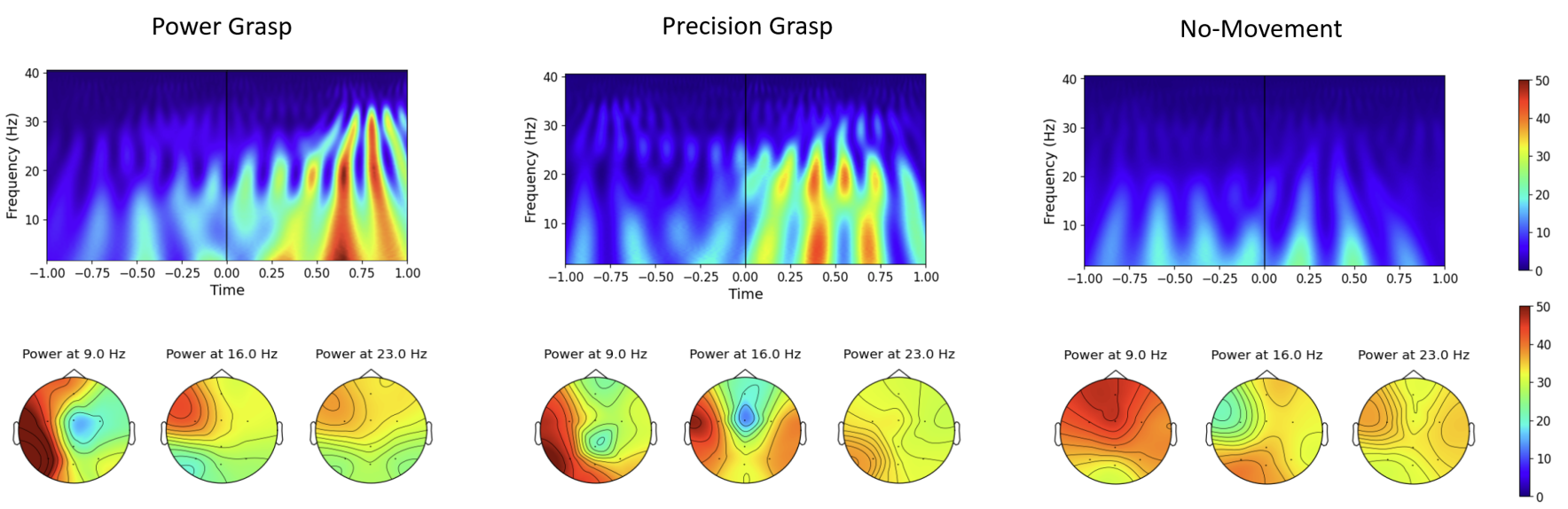}
\caption{The power of wavelet coefficients in EEG during hand-grasping tasks, with time-frequency plots for different conditions, and corresponding topographic maps focusing on 300ms post-movement across various frequencies.}
\label{wavelet_power}
\end{figure*}

\subsection{Classification}

In our study, we conducted a detailed analysis to classify hand grasp types using EEG data, employing advanced machine learning techniques and comprehensive feature extraction methods. Initially, we focused on extracting statistical features from the amplitude of continuous complex Morlet wavelet coefficients. Considering 8 EEG channels and 4 frequency bands, we extracted four statistical features—mean, variance, skewness, and kurtosis—from each band-channel combination, resulting in 128 features for each instance. For the classification of hand grasp types, we utilized four machine learning models: SVM, RF, XGBoost, and LDA. These models were applied to both multiclass and binary classification tasks, encompassing various hand-grasping conditions. To evaluate the models' performance, we adopted a 5-fold cross-validation approach, which provided a robust estimate of the models' generalization capabilities by reporting the average performance metrics across all folds \cite{kohavi1995study}.  Further, we conducted a permutation feature importance analysis on the best-performing model in multiclass classification (XGBoost) to identify key features for accurate classification. This involved randomly shuffling each feature and noting the impact on model performance. The results were visualized using boxplots and topographic maps, which effectively illustrated the spatial distribution of feature importance across brain regions. 


\section{Results}

\subsection{Wavelet Analysis}

Figure 2 combines time-frequency plots and topographic maps to analyze the C3 channel (over the motor cortex) during grasping tasks. The time-frequency plots show a surge in alpha and beta band activity, starting around 300ms after the movement onset, especially during power grasps. The topographic maps at 300ms post-movement confirm this, with peak activity at 9 Hz and a decline at higher frequencies, illustrating a frequency-dependent neural response. Across both analyses, grasping notably increases brain activity around the motor-centric C3 channel compared to no movement.

\subsection{Classification}

The classification results in Table 1 show that all four machine learning algorithms achieved accuracies above the 33.33\% chance level for three-class classification, indicating the effectiveness of the methods used to classify hand-grasping actions. XGBoost was the most accurate algorithm with an average of 85.16\%, closely followed by Random Forest with 83.10\%.

\begin{table}[ht!] 
\centering  
\caption{Accuracies for multiclass classification}
\label{tab:results_part2}
\begin{tabular}{lcccc}
\toprule
Subjects & SVM (\%) & RF (\%) & XGBoost (\%) & LDA (\%) \\
\midrule
S1 & 68.00 & 83.11 & 82.44 & 71.11 \\
S2 & 68.81 & 81.28 & 84.49 & 64.17 \\
S3 & 70.00 & 82.00 & 86.89 & 71.22 \\
S4 & 72.67 & 84.44 & 87.56 & 68.89 \\
S5 & 76.67 & 84.67 & 84.44 & 71.33 \\
\midrule
Mean & \textit{71.23} & \textit{83.10} & \textit{85.16} & \textit{69.34} \\
STD & \textit{3.51} & \textit{1.48} & \textit{2.07} & \textit{3.06} \\
\bottomrule
\end{tabular}
\end{table}

\begin{table*}[ht!]
\centering  
\caption{Accuracies for binary classifications}
\label{tab:merged_results}
\begin{tabular}{lcccccccccccc}
\toprule
 & \multicolumn{4}{c}{No-Movement vs Power (\%)} & \multicolumn{4}{c}{No-Movement vs Precision (\%)} & \multicolumn{4}{c}{Power vs Precision (\%)} \\
\cmidrule(r){2-5} \cmidrule(lr){6-9} \cmidrule(l){10-13}
Subjects & SVM & RF & XGBoost & LDA & SVM & RF & XGBoost & LDA & SVM & RF & XGBoost & LDA \\
\midrule
S1 & 94.00 & 90.33 & 91.00 & 84.33 & 95.00 & 95.00 & 93.00 & 87.00 & 83.67 & 83.67 & 83.67 & 74.33 \\
S2 & 95.19 & 93.04 & 96.52 & 85.29 & 93.85 & 95.99 & 96.25 & 84.23 & 86.36 & 85.83 & 85.03 & 68.44 \\
S3 & 96.33 & 95.33 & 94.67 & 91.33 & 96.67 & 95.33 & 93.67 & 90.00 & 89.67 & 90.33 & 93.67 & 69.00 \\
S4 & 95.67 & 95.33 & 96.00 & 92.67 & 92.67 & 94.67 & 92.33 & 89.67 & 84.33 & 88.33 & 90.33 & 67.67 \\
S5 & 95.67 & 97.67 & 96.67 & 94.00 & 95.00 & 96.00 & 92.33 & 90.33 & 86.33 & 85.33 & 87.67 & 71.00 \\
\midrule
Mean & 95.37 & 94.34 & 94.97 & 89.52 & 94.64 & 95.40 & 93.52 & 88.25 & 86.07 & 86.70 & 88.07 & 70.08 \\
STD & 0.87 & 2.78 & 2.36 & 4.42 & 1.49 & 0.59 & 1.63 & 2.60 & 2.34 & 2.63 & 4.04 & 2.67 \\
\bottomrule
\end{tabular}
\end{table*}

Table 2, encompassing results for various binary classifications, reveals significant findings. In the No-Movement vs Power Grasp classification, all models surpassed the 50\% chance level, with SVM leading at 95.37\% accuracy and XGBoost closely following. In the No-Movement vs Precision Grasp category, Random Forest emerged as the top performer with an accuracy of 95.40\%, with SVM also showing strong results at 94.64\%. The classification between Power Grasp and Precision Grasp indicated a general decrease in model accuracies, reflecting the complexity of distinguishing between these grasp types compared to no-movement vs grasp conditions. Here, LDA showed a notable decrease in performance, averaging 70.08\% accuracy. However, XGBoost maintained high efficacy, leading with 88.07\% accuracy, indicating its robustness in handling complex EEG data classification challenges.

Figure 3 merges analyses of feature importance with their topographical distribution across brain regions, illustrating the significance of specific EEG features in classifying hand-grasping actions. Figure 3a identifies 'C3 alpha kurtosis' as the most critical feature for classifying hand grasps, with 'C4 beta skewness' and 'PO7 beta mean' also significant, indicating the relevance of alpha and beta bands. Cerebral, and parietal, regions are shown to be important in motor functions. Figure 3b, through topographic maps, visually highlights the alpha and beta band features' importance across the brain, with the beta band features, particularly around the C3 and C4 channels, marked as crucial for motor planning.

\begin{figure*}[ht!] 
\centering
\includegraphics[width=7in]{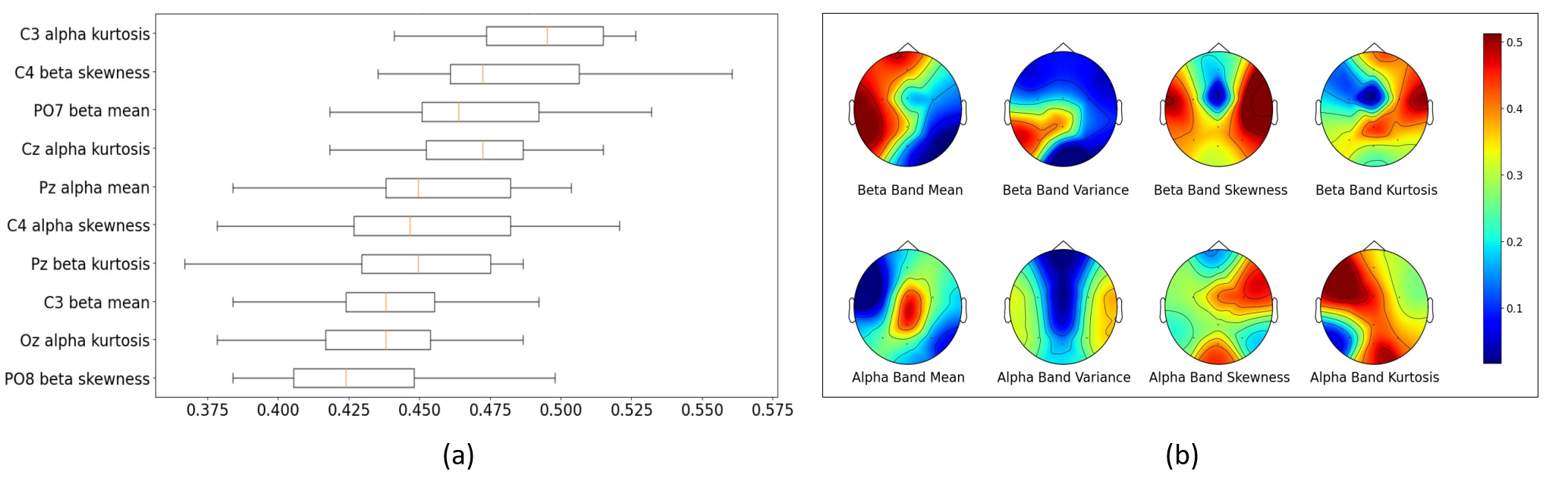}
\caption{(a) Boxplot illustrating the permutation feature importance scores for the top 10 EEG features in the classification of hand-grasping actions. (b) Topographic maps showing the spatial distribution of feature importance across the brain at different frequency bands.}
\label{feature_importance}

\end{figure*}


\section{Discussion}

In our comprehensive wavelet analysis,  the results illustrated in Figure 2 showed a notable rise in brain activity around 300ms after movement onset, especially in the alpha and beta bands, linked with the neural dynamics of grasping. The power grasp condition generated more wavelet power than precision grasp, confirming earlier research \cite{iturrate2018human}. Further spatial analysis through topographic maps indicated that after 300ms post-movement onset, there was a marked increase in power, especially in the 9Hz (alpha band) and 16Hz (beta band) frequencies, predominantly around the C3 channel. This channel's involvement is well-documented in its association with motor movements \cite{li2018combining}, and our findings corroborate this relationship, reflecting the localized neural activity linked to the hand-grasping actions.

The classification results, detailed in Tables 1 and 2, showcase our success in applying statistical wavelet features to differentiate between various grasp types. Our approach confirmed higher accuracy results in comparisons with similar studies using only temporal \cite{xu2021decoding, schwarz2017decoding}, spectral \cite{jochumsen2016detecting}, or combined features \cite{schwarz2019combining}, with the best accuracy for multiclass classification at 85.16\%, and for binary classification of grasp vs. no-movement at 95.40\%, as well as 88.07\% for binary classification between different grasp types. These results not only demonstrate the efficacy of our feature extraction and classification methods but also highlight the advantages of using a combination of temporal and spectral features through wavelet analysis.

Finally, the feature importance analysis in Figure 3 shows that key features for grasp classification are primarily in the brain's motor areas, especially within the alpha and beta bands. These findings align with existing neurophysiological research \cite{castiello2005neuroscience}, highlighting the importance of these regions and frequency bands in hand movement. Notably, features such as 'Cz alpha kurtosis' and 'C3 beta mean' in central and parietal areas are identified as highly influential. The 'Cz alpha kurtosis' reflects transitions between rest and motor execution, crucial for distinguishing between no-movement and active grasping, while the 'C3 beta mean' captures variations in motor planning and execution, essential for differentiating power from precision grasps.


\section{Conclusion}

This study advances EEG analysis for hand grasp classification by incorporating wavelet analysis, revealing increased brain activity in alpha and beta frequency bands during movements. The topographic maps underscore the critical role of the motor cortex, especially around the C3 and C4 channels. Utilizing advanced machine learning models, the study achieved high accuracy in classification, outperforming previous methods. Future studies should focus on applying these wavelet analysis insights into actual neuroprosthetic and BCI systems, diversifying motor tasks and participant demographics. Further advancements could include applying other wavelet functions and conducting long-term studies to verify the consistency of the EEG features.




\bibliographystyle{IEEEtran}
\bibliography{main}

\end{document}